\documentclass[preprint,aps,prd,showpacs,nofootinbib,superscriptaddress,12pt]{revtex4}%
\usepackage{graphicx}
\usepackage{amsmath}
\usepackage{amssymb}
\usepackage{bm}
\usepackage{epsfig}
\begin{document}
\title{\mbox{}\\[10pt]
Fragmentation contributions to $\bm{J/\psi}$ photoproduction
at HERA
}
\author{Geoffrey~T.~Bodwin}
\affiliation{High Energy Physics Division, Argonne National Laboratory,\\
9700 South Cass Avenue, Argonne, Illinois 60439, USA}
\author{Hee~Sok~Chung}
\affiliation{High Energy Physics Division, Argonne National Laboratory,\\
9700 South Cass Avenue, Argonne, Illinois 60439, USA}
\author{U-Rae~Kim}
\affiliation{Department of Physics, Korea University, Seoul 136-701, Korea}
\author{Jungil~Lee}
\affiliation{Department of Physics, Korea University, Seoul 136-701, Korea}

\date{\today}
\begin{abstract}
 We compute leading-power fragmentation corrections to $J/\psi$
photoproduction at DESY HERA, making use of the nonrelativistic QCD
factorization approach. Our calculations include parton production cross
sections through order $\alpha_s^3$, fragmentation functions though
order $\alpha_s^2$, and leading logarithms of the transverse momentum
divided by the charm-quark mass to all orders in $\alpha_s$. We find
that the leading-power fragmentation corrections, beyond those that are
included through next-to-leading order in $\alpha_s$, are small relative
to the fixed-order contributions through next-to-leading order in
$\alpha_s$. Consequently, an important discrepancy remains between the
experimental measurements of the $J/\psi$ photoproduction cross section
and predictions that make use of nonrelativistic-QCD long-distance
matrix elements that are extracted from the $J/\psi$ hadroproduction
cross-section and polarization data.
\end{abstract}
\maketitle

\section{Introduction}

According to the nonrelativistic QCD (NRQCD) factorization 
conjecture~\cite{Bodwin:1994jh}, 
the inclusive production cross section to produce a quarkonium $H$ in a 
collision of particles $A$ and $B$ can be written as 
\begin{equation}
d \sigma_{A+B \to H + X}= \sum_n d \sigma_{A+B \to Q \bar Q(n) + X} 
\langle  {\cal O}^H(n) \rangle,
\label{NRQCD-fact}
\end{equation}
where the $d \sigma_{A+B \to Q \bar Q(n)+X}$ are the short-distance
coefficients (SDCs), which are essentially the production cross
section of the heavy quark-antiquark pair $Q \bar Q(n)$ in a specific
color and angular momentum state $n$, and the $\langle {\cal O}^H(n)
\rangle$ are the corresponding NRQCD long-distance matrix elements
(LDMEs), which account for the evolution of the $Q \bar Q(n)$ pair into
the quarkonium.

The LDMEs have a known scaling with $v$, the heavy-quark velocity in the
quarkonium rest frame \cite{Bodwin:1994jh}. Usually, in heavy-quarkonium
production phenomenology, $v$ is considered to be a small parameter
($v^2\approx 0.3$ for the $J/\psi$), and the sum over $n$ in
Eq.~(\ref{NRQCD-fact}) is truncated at relative order $v^4$. For $H =
J/\psi$, the sum over $n$, truncated at order $v^4$, involves four $Q
\bar Q$ LDMEs: $\langle{\cal O}^{J/\psi}({}^3S_1^{[1]})\rangle$,
$\langle{\cal O}^{J/\psi}({}^3S_1^{[8]})\rangle$, $\langle{\cal
O}^{J/\psi}({}^1S_0^{[8]})\rangle$, and $\langle{\cal
O}^{J/\psi}({}^3P_J^{[8]})\rangle$, where the expressions in parentheses
give the color state of the $Q\bar Q$ pair (singlet or octet) and 
the angular-momentum state in spectroscopic notation. The
color-singlet LDME $\langle {\cal O}^{J/\psi} ({}^3S_1^{[1]}) \rangle$
can be measured in lattice QCD, can be determined from potential models,
or can be computed from the $J/\psi$ decay rate into lepton pairs. It is not
known how to calculate the color-octet LDMEs from first principles. As a
result, they are usually obtained by comparing Eq.~(\ref{NRQCD-fact}) to
measured cross sections.

SDCs for the $J/\psi$ production cross sections and polarizations have
been computed through next-to-leading order (NLO) in the strong-coupling
constant $\alpha_s$ for
hadroproduction \cite{Ma:2010jj,Butenschoen:2010rq,Ma:2010yw,Chao:2012iv,
Butenschoen:2012px,Gong:2012ug} and
photoproduction \cite{Butenschoen:2009zy,Butenschoen:2011ks,Butenschoen:2011yh}.
The LDMEs that are obtained by fitting the resulting cross-section and
polarization predictions to the experimental data vary considerably,
depending on the data that are used in the fits. It is possible to fit
the hadroproduction cross-section data
\cite{Acosta:2004yw,Chatrchyan:2011kc} and polarization data
\cite{Affolder:2000nn,Abulencia:2007us,Chatrchyan:2013cla}
simultaneously \cite{Chao:2012iv}, but the resulting LDMEs give a
prediction for the photoproduction cross section that overshoots the data by 
factors of 4--6 at the highest value of the quarkonium transverse momentum 
$p_T$ at which the cross section has been measured \cite{Butenschoen:2012qr}.
Alternatively, one can extract the LDMEs by comparing the NLO
predictions for the hadroproduction and photoproduction cross sections
with the experimental data \cite{Butenschoen:2012px}, but the resulting
LDMEs lead to predictions of large transverse polarization in
hadroproduction at large $p_T$ that are at odds with the experimental
data \cite{Butenschoen:2012px}.

Recently, it was found that fragmentation contributions at leading power
(LP) in $p_T$ that go beyond NLO in $\alpha_s$ are important in $J/\psi$
hadroproduction \cite{Bodwin:2014gia}. With the inclusion of these
contributions it was possible, for the first time, to use the LDMEs that
have been extracted from the hadroproduction cross sections alone to
obtain predictions for the $J/\psi$ polarization at large $p_T$ that are
near zero and are in agreement with the experimental data
\cite{Bodwin:2014gia}. However, these LDMEs, when combined with the NLO
SDCs for the photoproduction cross section, lead to a prediction that
also overshoots the HERA data from the H1 Collaboration
\cite{Adloff:2002ex,Aaron:2010gz} by about a factor of 8 in the 
highest $p_T$ bin for which the cross section has been measured.

Motivated by this discrepancy between theory and experiment and by the
large LP fragmentation contributions to $J/\psi$ hadroproduction beyond
NLO in $\alpha_s$, we compute in this paper LP fragmentation
contributions to $J/\psi$ photoproduction. Our approach is based on the
method that was described in Ref.~\cite{Bodwin:2014gia}.

The remainder of this paper is organized as follows. In 
Sec.~\ref{sec:LP-frag}, we briefly review the general method for 
computing the LP fragmentation contributions that was given in 
Ref.~\cite{Bodwin:2014gia}. Section~\ref{sec:comp} 
contains a description of the details of the computation of the 
LP fragmentation contribution to photoproduction. In 
Sec.~\ref{sec:numerical}, we present our numerical results. Finally, in 
Sec.~\ref{sec:summary}, we summarize and discuss our results.

\section{LP Fragmentation \label{sec:LP-frag}}

At large transverse momentum $p_T$, the contribution at LP in $p_T$ to 
the cross section to produce a $Q\bar Q$ pair in two-body collisions is 
given by \cite{Collins:1981uw, Nayak:2005rt}
\begin{equation}
d \sigma^{\rm LP}_{A+B \to Q \bar Q(n) + X}=\sum_i d\hat\sigma_{A+B\to i+X}
\otimes D_{i\to Q \bar Q(n)},
\label{sigma-LP-fact}
\end{equation}
where the $d\hat\sigma_{A+B\to i+X}$ are the inclusive parton
production cross sections (PPCSs) to produce a parton $i$, and the
$D_{i\to Q\bar Q(n)}$ are the fragmentation functions (FFs) for a parton
$i$ to fragment into a $Q\bar Q$ pair in color and angular-momentum
state $n$. At the parton level, before convolution with parton
distribution functions, the LP contribution to $d \sigma/d p_T^2$
depends asymptotically on $p_T$ as $1/p_T^4$.

In this paper, we consider gluon fragmentation and light-quark
fragmentation. The gluon FFs $D_{g\to Q\bar Q(n)}$ in
Eq.~(\ref{sigma-LP-fact}) are given for the ${}^1S_0^{[8]}$ channel at
order $\alpha_s^2$ (LO) in Refs.~\cite{Braaten:1996rp,Bodwin:2012xc},
for the ${}^3S_1^{[8]}$ channel at order $\alpha_s$ (LO) in
Ref.~\cite{Braaten:1994kd} and at order $\alpha_s^2$ (NLO) in
Refs.~\cite{Braaten:2000pc,Ma:2013yla}, and for the ${}^3P_J^{[8]}$
channels at order $\alpha_s^2$ (LO) in
Refs.~\cite{Braaten:1994kd,Bodwin:2012xc}. The light-quark FF $D_{q\to
Q\bar{Q}(n)}$ in the ${}^3S_1^{[8]}$ channel is given at order
$\alpha_s^2$ (LO) in
Refs.~\cite{Ma:1995vi,Ma:2013yla,hong-zhang,Bodwin:2014bia,Ma:2015yka}.
Light-quark fragmentation in the other color-octet channels vanishes
through order $\alpha_s^2$. In the color-singlet channel, gluon
fragmentation occurs at order $\alpha_s^3$
(LO)~\cite{Braaten:1995cj}. Light-quark fragmentation in the
color-singlet channel vanishes through order $\alpha_s^2$, but
charm-quark fragmentation occurs at order
$\alpha_s^2$~\cite{Bodwin:2014bia, Ma:2015yka}. As we will explain in
more detail below, an estimate of the size of the color-singlet
LP fragmentation contribution shows that it is negligible in comparison
with the overall theoretical uncertainties. Therefore, we will ignore
the color-singlet LP fragmentation contributions in our numerical
analyses.

The FFs depend on the factorization scale $\mu_f$. The dependence on
$\mu_f$ is governed by the Dokshitzer-Gribov-Lipatov-Altarelli-Parisi
(DGLAP) evolution equation~\cite{Gribov:1972ri, Lipatov:1974qm,
Dokshitzer:1977sg, Altarelli:1977zs}. At leading order in $\alpha_s$,
the DGLAP equation reads
\begin{equation}
\frac{d}{d \log \mu_f^2}               
\begin{pmatrix}     
D_S \\ D_g
\end{pmatrix}
=
\frac{\alpha_s(\mu_f)}{2 \pi}
\begin{pmatrix}
P_{qq} & 2 n_f P_{gq} \\
P_{qg} & P_{gg}
\end{pmatrix}
\otimes
\begin{pmatrix}
D_S \\ D_g
\end{pmatrix},
\label{eq:DGLAP}
\end{equation}
where $D_g = D_{g \to Q \bar Q(n)}$, $D_S = \sum_f [ D_{q_f \to Q \bar
Q(n)} + D_{\bar q_f \to Q \bar Q(n)}]$, $f$ is the light-quark or
light-antiquark flavor, the $P_{ij}$ are the splitting functions for
the FFs, and $n_f$ is the number of active light-quark flavors. In order
to match what was done in the NLO calculations of
Refs.~\cite{Butenschoen:2009zy,Butenschoen:2011ks}, we take $n_f=3$.
We resum the leading logarithms of $p_T/m_c$ by solving
Eq.~(\ref{eq:DGLAP}) to evolve the FFs from $\mu_f = 2 m_c$ to $\mu_f =
m_T = \sqrt{p_T^2+4 m_c^2}$.

Following Ref.~\cite{Bodwin:2014gia}, we combine the LO-plus-NLO SDCs 
with the LP fragmentation contributions according to the formula 
\begin{equation}
\frac{d\sigma^{\rm LP+NLO}}{dp_T}
=\frac{d\sigma^{\textrm{LP}}}{dp_T}
-\frac{d\sigma^{\textrm{LP}}_{\rm NLO}}{dp_T}
+\frac{d\sigma_{\rm NLO}}{dp_T}.
\end{equation}
Here, ${d\sigma^{\textrm{LP}}}/{dp_T}$ is the DGLAP-evolved LP
fragmentation contribution, ${d\sigma_{\rm NLO}}/{dp_T}$ is the
contribution that arises from the LO-plus-NLO SDCs, and
${d\sigma^{\textrm{LP}}_{\rm NLO}}/{dp_T}$ is the contribution that
is contained in both ${d\sigma^{\textrm{LP}}}/{dp_T}$ and
${d\sigma_{\rm NLO}}/{dp_T}$.

\section{Computation of LP SDCs\label{sec:comp}}

In photoproduction at HERA, the incoming electron or positron emits a
virtual photon that is nearly on mass shell and that subsequently
interacts with the incoming proton. There are two types of
photon-induced processes that contribute to photoproduction cross
sections. The first is the {\it direct} process, in which the virtual
photon interacts with a parton in the proton electromagnetically. The
second is the {\it resolved} process, in which the virtual photon emits
a parton, which then interacts strongly with a parton in the proton.
The probability for the photon to emit a parton is given by the parton
distribution function (PDF) of the photon.

We compute the PPCSs $d \hat \sigma_{e+p \to i + X}$ for direct and
resolved photoproduction to NLO accuracy in $\alpha_s$ by making use of
the EPHOX Fortran code~\cite{Fontannaz:2001ek, Fontannaz:2001nq,
Fontannaz:2002nu, Fontannaz:2003yn}. We use the AFG04\_BF photon PDFs
and the CTEQ6M proton PDFs at scale $m_T$. We carry out the computation
using the same kinematics and cuts as were used by the H1 Collaboration
in their most recent cross-section measurements
~\cite{Aaron:2010gz}. The center-of-momentum energy of the $e p$ system is
$\sqrt{s} = 319$ GeV. The cuts on the $\gamma p$-invariant
mass $W = \sqrt{(p_\gamma+p_p)^2}$ and elasticity $z = p_{J/\psi} \cdot
p_p/p_\gamma \cdot p_p$ are given by $60\textrm{ GeV}<W<240\textrm{
GeV}$ and $0.3 < z < 0.9$. Here, $p_\gamma$, $p_p$, and $p_{J/\psi}$ are
the momenta of the photon, proton, and $J/\psi$, respectively. There is
also a cut on the invariant mass of the virtual photon $Q^2$, which is
$Q^2 < Q_{\rm max}^2 = 2.5$ GeV${}^2$. The photon flux is
calculated in EPHOX by making use of  the Weizs\"acker-Williams formula
for quasireal photons:
\begin{equation}
f_{\gamma/e} (x) 
= \frac{\alpha}{2 \pi} 
\left[ 
\frac{1+ (1-x)^2}{x} \log \frac{Q^2_{\rm max} (1-x)}{m_e^2 x^2}
- \frac{2 (1-x)}{x}
\right].
\end{equation}
Here, $x = E_\gamma/E_e$, where $E_\gamma$ and $E_e$ are the energy of the
photon and the electron, respectively; $\alpha$ is the quantum
electrodynamics coupling constant; and $m_e$ is the electron mass.

\section{Numerical Results\label{sec:numerical}}

\begin{figure}
\epsfig{file=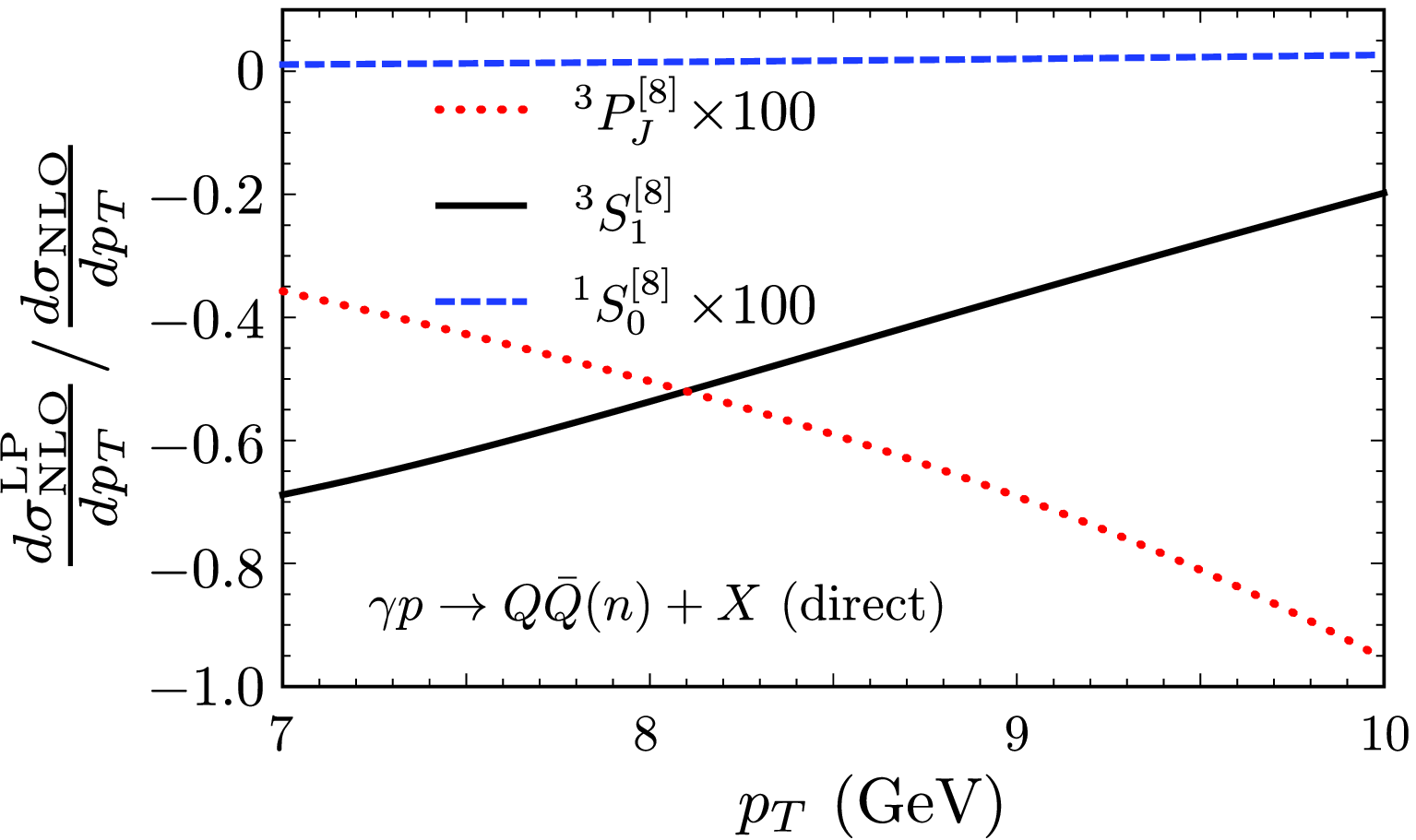,width=15cm}
\caption{\label{fig:NLO-frag-direct}%
The ratio
$(d\sigma^{\textrm{LP}}_{\rm NLO}/dp_T) / (d\sigma_{\rm NLO}/dp_T)$
for the ${}^1S_0^{[8]}$, ${}^3P_J^{[8]}$, and ${}^3S_1^{[8]}$ channels
in the direct process $\gamma p \to J/\psi + X$. 
}
\end{figure}

\begin{figure}
\epsfig{file=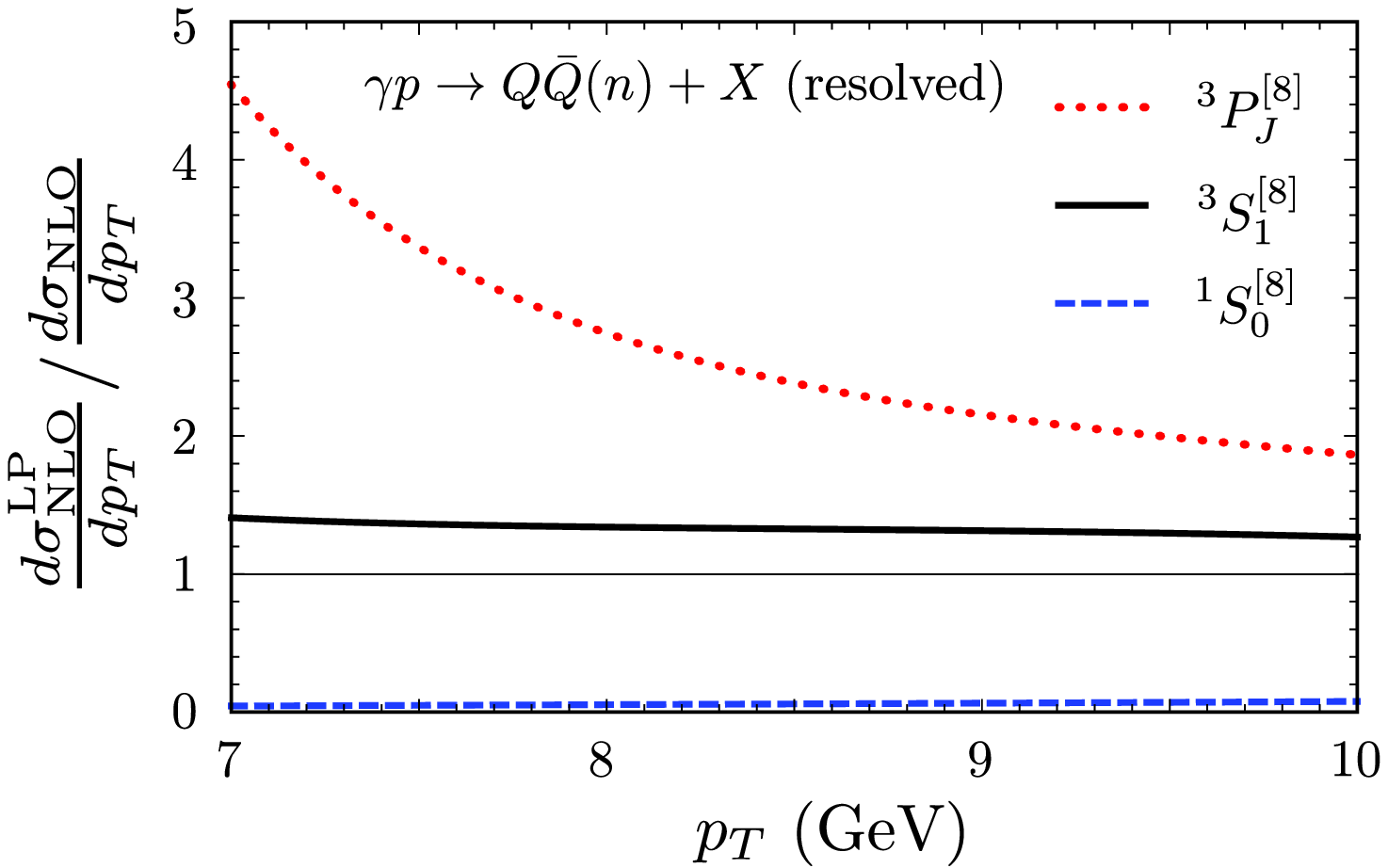,width=15cm}
\caption{\label{fig:NLO-frag-resolved}%
The ratio
$(d\sigma^{\textrm{LP}}_{\rm NLO}/dp_T) / (d\sigma_{\rm NLO}/dp_T)$
for the ${}^1S_0^{[8]}$, ${}^3P_J^{[8]}$, and ${}^3S_1^{[8]}$ channels
in the resolved process $\gamma p \to J/\psi + X$.
}
\end{figure}

In Fig.~\ref{fig:NLO-frag-direct}, we compare our results for
$d\sigma^{\textrm{LP}}_{\rm NLO}/dp_T$ for the direct process with the
NLO calculation for the direct process. In
Fig.~\ref{fig:NLO-frag-resolved}, we make a similar comparison for the
resolved process. Here, and throughout the remainder of this paper, we
make use of the SDCs through NLO that were computed in
Refs.~\cite{Butenschoen:2009zy,Butenschoen:2011ks,Butenschoen:2012qr}.

For the direct process, LP fragmentation has a sizable contribution only
for the ${}^3S_1^{[8]}$ channel. In the ${}^1S_0^{[8]}$ channel,
$d\sigma_{\rm NLO}^{\rm LP}/dp_T$ is very small in comparison to
$d\sigma_{\rm NLO}/dp_T$ because the PPCSs involve only
light-quark-initiated processes, which have much smaller partonic fluxes
than do gluon-initiated processes, and because the FFs 
contain no distributions that emphasize the region near $z=1$. In the
${}^3P_J^{[8]}$ channels, $d\sigma_{\rm NLO}^{\rm LP}/dp_T$ is also
small in comparison to $d\sigma_{\rm NLO}/dp_T$ (less than 1\% for $p_T
\le 10$~GeV) because the PPCSs involve only light-quark-initiated
processes. In the ${}^3P_J^{[8]}$ channels, $d\sigma_{\rm NLO}^{\rm
LP}/dp_T$ is opposite in sign to the non-LP part of $d\sigma_{\rm
NLO}/dp_T$ and, as expected, grows in magnitude relative to the non-LP
part of $d\sigma_{\rm NLO}/dp_T$ as $p_T$ increases. However, even at
$p_T=10$~GeV, $d\sigma_{\rm NLO}^{\rm LP}/dp_T$ is much smaller than the
non-LP part of $d\sigma_{\rm NLO}/dp_T$. Consequently, the ratio
$(d\sigma_{\rm NLO}^{\rm LP}/dp_T)/(d\sigma_{\rm NLO}/dp_T)$ becomes
increasingly negative as $p_T$ increases. At very large values of $p_T$,
at which the magnitude of $d\sigma_{\rm NLO}^{\rm LP}/dp_T$ becomes
comparable to or larger than the magnitude of the non-LP part of
$d\sigma_{\rm NLO}/dp_T$, we would expect this ratio to change sign
discontinuously and to approach unity. We do not show results for the
${}^3S_1^{[1]}$ channel. We defer the discussion of that channel until
we discuss the sum of the direct and resolved contributions to the cross
section.

For the resolved process, LP fragmentation in the ${}^3S_1^{[8]}$ and
${}^3P_J^{[8]}$ channels approaches the NLO calculation as $p_T$ rises.
In the ${}^1S_0^{[8]}$ channel, the LP fragmentation contribution is
small because, unlike the FFs for the ${}^3S_1^{[8]}$ and ${}^3P_J^{[8]}$ channels,
the FF for the ${}^1S_0^{[8]}$ channel does not involve 
distributions that emphasize the region near $z=1$.
Again, we defer the discussion 
of the ${}^3S_1^{[1]}$ channel until we discuss the sum of the direct 
and resolved contributions to the cross section.

\begin{figure}
\epsfig{file=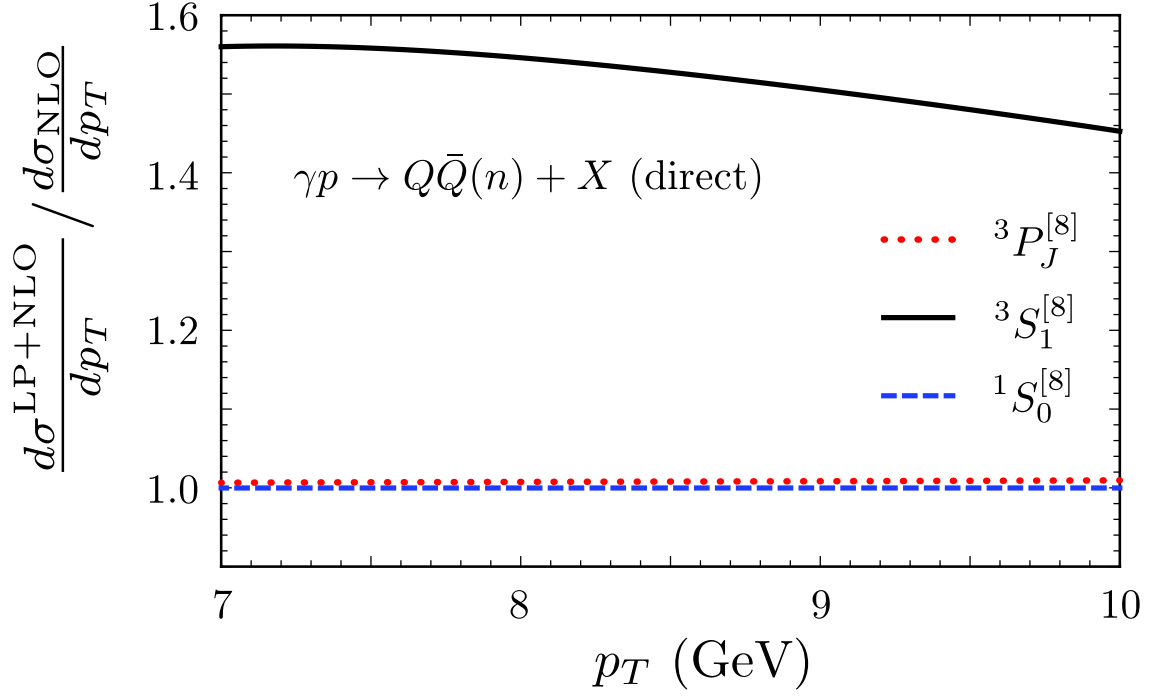,width=15cm}
\caption{\label{fig:LPNLO-direct}%
The ratio
$(d\sigma^{\rm LP+NLO}/dp_T) / (d\sigma_{\rm NLO}/dp_T)$
for the ${}^1S_0^{[8]}$, ${}^3P_J^{[8]}$, and ${}^3S_1^{[8]}$ channels
in the direct process $\gamma p \to J/\psi + X$. 
}
\end{figure}

\begin{figure}
\epsfig{file=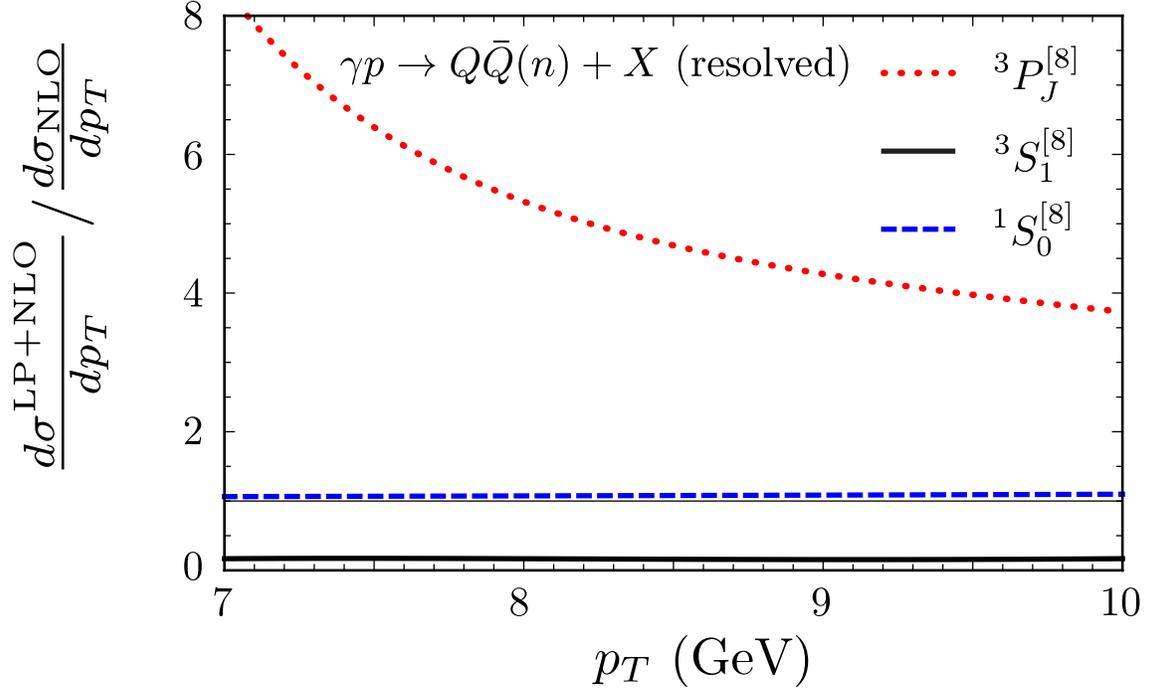,width=15cm}
\caption{\label{fig:LPNLO-resolved}%
The ratio
$(d\sigma^{\rm LP+NLO}/dp_T) / (d\sigma_{\rm NLO}/dp_T)$
for the ${}^1S_0^{[8]}$, ${}^3P_J^{[8]}$, and ${}^3S_1^{[8]}$ channels
in the resolved process $\gamma p \to J/\psi + X$.
}
\end{figure}

In Figs.~\ref{fig:LPNLO-direct} and~\ref{fig:LPNLO-resolved}, we compare
$d\sigma^{\textrm{LP+NLO}}/dp_T$ with $d\sigma_{\rm NLO}/dp_T$ in each
channel for the direct and the resolved process, respectively. For the
direct process, $d\sigma^{\textrm{LP+NLO}}/dp_T$ is larger than
$d\sigma_{\rm NLO}/dp_T$ in the ${}^3S_1^{[8]}$ channel, while in the
${}^3P_J^{[8]}$ and ${}^1S_0^{[8]}$ channels, the differences between
$d\sigma^{\textrm{LP+NLO}}/dp_T$ and $d\sigma_{\rm NLO}/dp_T$ are less
than 1\% and less than 0.02\%, respectively. For the resolved
process, the difference between $d\sigma^{\textrm{LP+NLO}}/dp_T$ and
$d\sigma_{\rm NLO}/dp_T$ is substantial in the ${}^3S_1^{[8]}$ and
${}^3P_J^{[8]}$ channels. In the ${}^1S_0^{[8]}$ channel,
$d\sigma^{\textrm{LP+NLO}}/dp_T$ is larger than $d\sigma_{\rm NLO}/dp_T$
by only 7\% at $p_T = 7$~GeV and by only 10\% at $p_T = 10$~GeV.

\begin{figure}
\epsfig{file=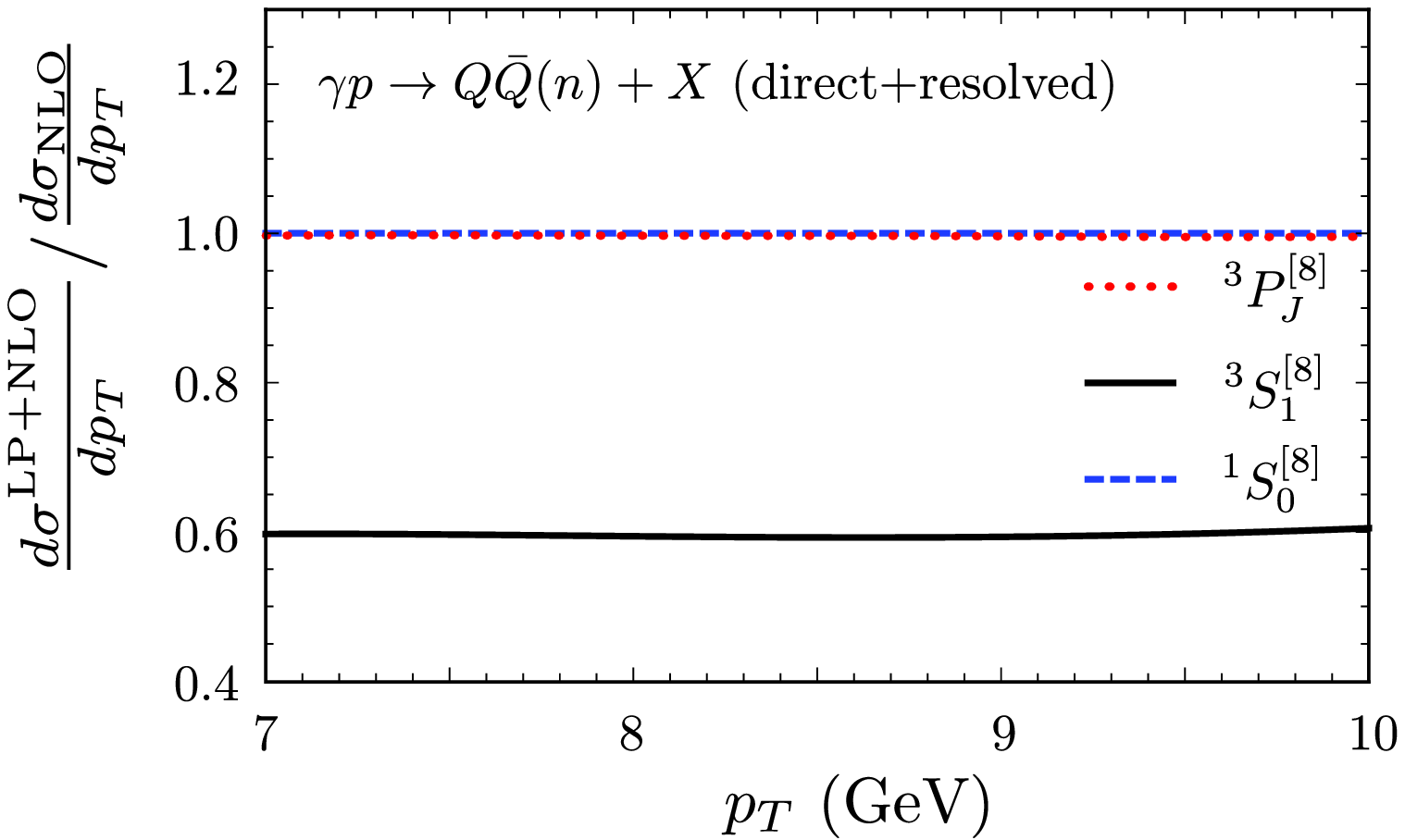,width=15cm}
\caption{\label{fig:LPNLO}%
The ratio $(d\sigma^{\rm LP+NLO}/dp_T) / (d\sigma_{\rm NLO}/dp_T)$ for
the ${}^1S_0^{[8]}$, ${}^3P_J^{[8]}$, and ${}^3S_1^{[8]}$ channels in
the sum of the direct and resolved processes $\gamma p \to J/\psi + X$.
}
\end{figure}

In Fig.~\ref{fig:LPNLO}, we compare, for each channel, the sum of the
direct and resolved contributions to $d\sigma^{\textrm{LP+NLO}}/dp_T$
with the sum of the direct and resolved contributions to $d\sigma_{\rm
NLO}/dp_T$. 
In the ${}^3S_1^{[8]}$ channel, $d\sigma^{\textrm{LP+NLO}}/dp_T$ is
smaller than $d\sigma_{\rm NLO}/dp_T$ by 40\% for $p_T \le 10$~GeV. On
the other hand, in the ${}^1S_0^{[8]}$ and ${}^3P_J^{[8]}$ channels, the
differences between $d\sigma^{\textrm{LP+NLO}}/dp_T$ and $d\sigma_{\rm
NLO}/dp_T$ are negligible. The reason for this is that, owing to the
experimental kinematic constraints on $W$ and $z$, the direct process
dominates over the resolved process in these channels, and the
differences between $d\sigma^{\textrm{LP+NLO}}/dp_T$ and $d\sigma_{\rm
NLO}/dp_T$ are small in direct production in these
channels~\cite{Butenschoen:2012qr}. We do not show the contribution
of the ${}^3S_1^{[1]}$ channel. We have made a rough estimate of the
size of the LP fragmentation correction in that channel by making use of
the gluon and charm-quark FFs at LO in $\alpha_s$. We estimate that, in
the ${}^3S_1^{[1]}$ channel, the LP fragmentation contributions for
$9$~GeV$\le p_T\le 10$~GeV are less than $5\%$ of the sum of the
color-singlet direct and resolved contributions.

Finally, we can examine the effect of the additional LP fragmentation
contributions on the complete cross section. As we have seen, the
additional LP fragmentation contribution is a sizable fraction of the
rate only for the ${}^3S_1^{[8]}$ channel. However, the SDC through
NLO for the ${}^3S_1^{[8]}$ channel is much smaller than the SDCs
through NLO for the ${}^1S_0^{[8]}$ and ${}^3P_J^{[8]}$ channels.
Hence, the correction from the additional LP fragmentation contributions
to the $J/\psi$ photoproduction cross section is very small. For
example, if we use the color-octet LDMEs that were obtained in
Ref.~\cite{Bodwin:2014gia} by fitting the LP+NLO SDCs to
CMS~\cite{Chatrchyan:2011kc} and CDF data~\cite{Acosta:2004yw} and if we
use the color-singlet LDME that was obtained in
Ref.~\cite{Bodwin:2007fz} by making use of the $J/\psi$ leptonic
decay-rate data, then we find that the difference between
$d\sigma^{\textrm{LP+NLO}}/dp_T$ and $d\sigma_{\rm NLO}/dp_T$ for
$J/\psi$ photoproduction is less than 1\% for $p_T$ between 7~GeV and
10~GeV. If we use the LDMEs from Ref.~\cite{Butenschoen:2012qr}, then
the difference between $d\sigma^{\textrm{LP+NLO}}/dp_T$ and
$d\sigma_{\rm NLO}/dp_T$ is less than 3\% for the same $p_T$ range.

\begin{figure}
\epsfig{file=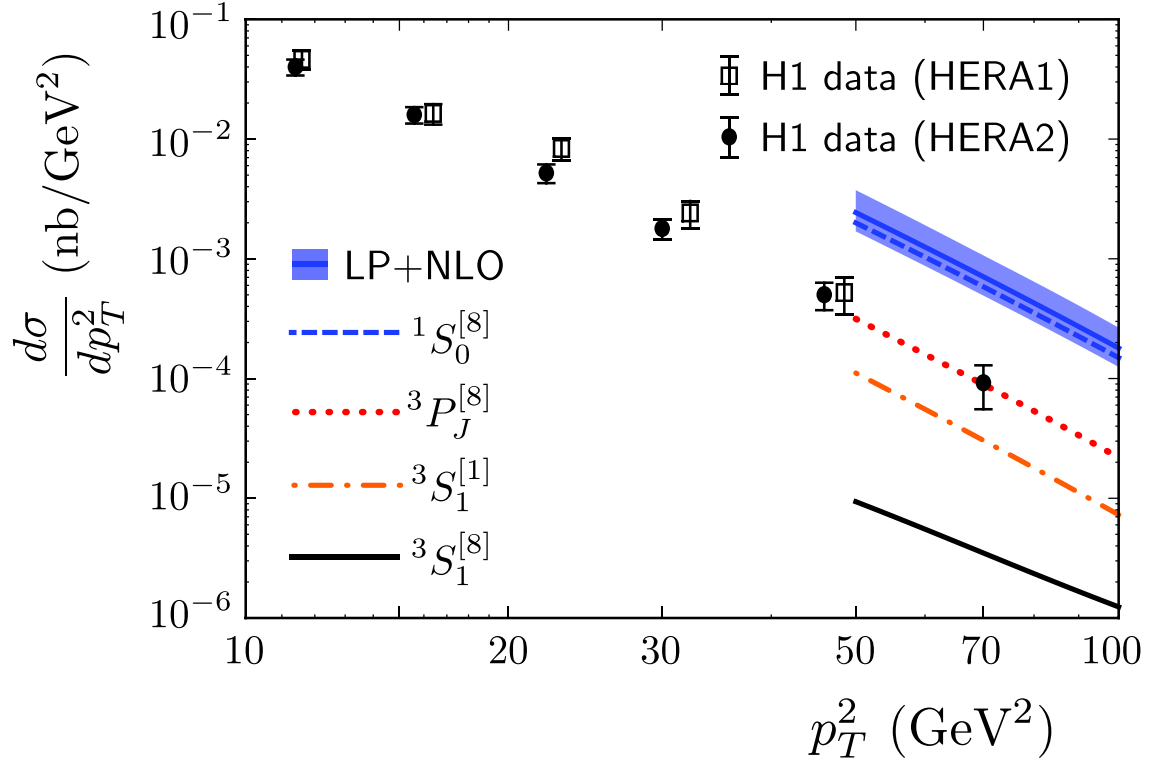,width=15cm}
\caption{\label{fig:crosssection}%
LP$+$NLO predictions for the $J/\psi$ differential cross section at HERA 
compared with the H1 data~\cite{Adloff:2002ex, Aaron:2010gz}.
}
\end{figure}

In Fig.~\ref{fig:crosssection}, we compare the  H1
data~\cite{Adloff:2002ex, Aaron:2010gz} for the $J/\psi$ photoproduction
cross section with the LP$+$NLO prediction that results from the use of
the color-octet LDMEs of Ref.~\cite{Bodwin:2014gia} and the
color-singlet LDME of Ref.~\cite{Bodwin:2007fz}. The uncertainty band in
the LP$+$NLO prediction comes from the uncertainties in the LDMEs,
combined in quadrature with the uncertainties in the SDCs that were
obtained in Ref.~\cite{Butenschoen:2012qr} by varying the
factorization scale $\mu_f$ between $2m_T$ and $m_T/2$. As we have
already mentioned, the additional LP fragmentation contributions do not
significantly change the prediction from that of the NLO calculation.
Consequently, the LP$+$NLO prediction overshoots the data by about a
factor of 8 at the highest value of $p_T$ at which the 
photoproduction cross section has been measured.

In order to suppress next-to-leading-power (NLP) contributions, we have
compared the theoretical predictions with data only for values of $p_T$
that are considerably larger than $m_{J/\psi}$. In
Ref.~\cite{Faccioli:2014cqa}, it was suggested that a criterion $p_T>
3 m_{J/\psi}$ be used in comparing data with theory. In
Ref.~\cite{Bodwin:2014gia}, a criterion $p_T\ge 10$~GeV was used in
fitting the predictions to the data. The highest value of $p_T$ at which
the photoproduction cross section has been measured falls slightly
short of these criteria. It is possible that NLP corrections and/or
power-suppressed violations of NRQCD factorization could account for
some of the differences between the LP$+$NLO prediction and the
experimental data. However, the data and the prediction do not seem to
be trending toward each other as $p_T$ increases.

The color-octet LDMEs of
Ref.~\cite{Bodwin:2014gia} were extracted from the prompt $J/\psi$
production data, which include feeddown from the $\chi_{cJ}$ and $\psi(2S)$
states. We would expect the corrections to the LDMEs from the removal of the 
feeddown contributions to prompt hadroproduction to be of order $-30\%$
\cite{Abe:1997yz}, and we would expect the corrections from the
inclusion of feeddown contributions to prompt photoproduction to be of
order $+15\%$ \cite{Aaron:2010gz}. This still leaves a substantial
discrepancy between the data and theoretical predictions.

\section{Summary and Discussion\label{sec:summary}}

In this paper, we have computed additional leading-power (LP) fragmentation
contributions to $J/\psi$ photoproduction at HERA that go beyond the
existing fixed-order calculations through NLO in $\alpha_s$. Our
computation has made use of parton production cross sections through NLO
in $\alpha_s$, as implemented in EPHOX, for both direct and resolved
photoproduction. We have included gluon and light-quark fragmentation
processes and have used the current state-of-the-art fragmentation
functions for the various color-octet $Q\bar Q$ channels. For gluon
fragmentation, the fragmentation function for the ${}^3S_1^{[8]}$
channel is available through NLO in $\alpha_s$, and the fragmentation
functions for the ${}^1S_0^{[8]}$ and ${}^3P_J^{[8]}$ channels are
available at LO in $\alpha_s$. For light-quark fragmentation, only the
LO fragmentation function for  the ${}^3S_1^{[8]}$ channel is available.
In addition to making use of the available fixed-order fragmentation
functions, we have resummed leading logarithms of $p_T^2/m_c^2$ to all
orders in $\alpha_s$ for the channels in which fragmentation functions
are available. We have estimated the LP fragmentation corrections in the
${}^3S_1^{[1]}$ channel by making use of the gluon and charm-quark
fragmentation functions at LO in $\alpha_s$.

We find that the additional LP fragmentation contributions are important,
relative to the fixed-order contributions through NLO in $\alpha_s$, only
in the ${}^3S_1^{[8]}$ channel in direct production and only in the
${}^3P_J^{[8]}$ channel in resolved production. However, the fixed-order
contributions from direct production in the ${}^3S_1^{[8]}$ channel are
themselves small, and the contributions from resolved production
are small in the ${}^3P_J^{[8]}$ channel (and the ${}^1S_0^{[8]}$ 
channel) for the kinematics and cuts of the H1 cross-section
measurement. Furthermore, all of the additional LP fragmentation
contributions in the ${}^3S_1^{[1]}$ channel are negligible compared to
the theoretical uncertainty in the sum of the contributions of all of
the channels. Consequently, the additional LP fragmentation
contributions that we have computed have little effect on the cross-section 
prediction.

If one predicts the photoproduction cross section by using NRQCD long-distance matrix elements
that are consistent with both the hadroproduction cross-section and
polarization data, then there is a sizable discrepancy between theory
and experiment at the highest values of $p_T$ at which the
photoproduction cross section has been measured. If NRQCD
factorization is correct, then one would expect it to hold at values of
$p_T$ that are considerably larger than $m_{J/\psi}$. As we have
mentioned, the highest value of $p_T$ at which the photoproduction cross
section has been measured is not very large in comparison with
$m_{J/\psi}$. Hence, it is possible that NLP corrections and/or
power-suppressed violations of NRQCD factorization could account for
some of the discrepancy between theory and the experimental data.
However, the shapes of the data and the LP$+$NLO prediction versus $p_T$
do not suggest a resolution of the discrepancy at larger values of
$p_T$. Hence, the discrepancy between theory and experiment in
photoproduction of the $J/\psi$ seems to challenge the validity of
NRQCD factorization.

The calculations in this paper include some, but not all, of the
nonlogarithmic LP fragmentation contributions at next-to-next-to-leading
order (NNLO) in $\alpha_s$ (order $\alpha_s^5$). A complete calculation
of the nonlogarithmic LP fragmentation contributions through NNLO in
$\alpha_s$ would require the calculation of additional QCD corrections to fragmentation
functions and parton production cross sections. In the case of gluon
fragmentation, which dominates in hadroproduction, a complete NNLO
calculation in the ${}^1S_0^{[8]}$ and ${}^3P_J^{[8]}$ channels would
require the use of fragmentation functions through NLO and parton
production cross sections through NLO, and a complete calculation in the
 ${}^3S_1^{[8]}$ channel would require the use of the fragmentation
function through NNLO and parton production cross sections through
NNLO. In the case of light-quark fragmentation, a complete NNLO
calculation in the ${}^1S_0^{[8]}$ and ${}^3P_J^{[8]}$ channels would
require the use of fragmentation functions through NLO and the LO parton
production cross sections, and a complete calculation in the
${}^3S_1^{[8]}$ channel would require the use of the fragmentation
function through NLO and parton production cross sections through NLO.
As we have mentioned, the fragmentation functions are all known only at
LO, except in the case of gluon fragmentation in the ${}^3S_1^{[8]}$
channel, in which case the fragmentation function is known through NLO.
Parton production cross sections are publicly available through NLO.

A complete NNLO calculation of the LP fragmentation contributions might
be important for hadroproduction cross sections and polarizations, and,
hence, could affect the extractions of the long-distance matrix elements
from the hadroproduction data. However, given the small sizes of the
additional LP fragmentation contributions that we have found in this
paper, it seems unlikely that these further LP fragmentation
contributions would be important for photoproduction.

\begin{acknowledgments}
We thank Mathias Butensch\"on and Bernd Kniehl for supplying
detailed numerical results from their NLO calculations of the
photoproduction SDCs. We also thank Michel Fontannaz for advice
regarding the use of the EPHOX code and Michel Fontannaz and
Jean-Philippe Guillet for providing a version of the EPHOX code that
contains the AFG04 photon parton distributions. The work of G.T.B.\ and
H.S.C.\ is supported by the U.S.\ Department of Energy, Division of High
Energy Physics, under Contract No.\ DE-AC02-06CH11357. The work of U-R.\,K.\
is supported by the National Research Foundation of Korea under Contract
No.\ NRF-2012R1A1A2008983. 
J.L.\ and U-R.K.\ thank APCTP for its hospitality through the CAT program.
The submitted manuscript has been created in
part by UChicago Argonne, LLC, Operator of Argonne National Laboratory.
Argonne, a U.S.\ Department of Energy Office of Science laboratory, is
operated under Contract No.\ DE-AC02-06CH11357. The U.S. Government
retains for itself, and others acting on its behalf, a paid-up
nonexclusive, irrevocable worldwide license in said article to
reproduce, prepare derivative works, distribute copies to the public,
and perform publicly and display publicly, by or on behalf of the
Government.

\end{acknowledgments}


\end{document}